\documentclass[conference]{IEEEtran}
\usepackage{cite}
\usepackage{amsmath,amssymb,amsfonts}
\usepackage{algorithmic}
\usepackage{graphicx}
\usepackage{textcomp}
\usepackage{xcolor}
\usepackage{url}
\def\BibTeX{{\rm B\kern-.05em{\sc i\kern-.025em b}\kern-.08em
    T\kern-.1667em\lower.7ex\hbox{E}\kern-.125emX}}
    
\begin{document}

\title{From Lemons to Peaches: Improving Security\\ ROI through Security Chaos Engineering}

\author{\IEEEauthorblockN{Kelly Shortridge}
\IEEEauthorblockA{\textit{Product Technology, Fastly}\\
New York, NY \\
0000-0002-5580-3164}
}

\maketitle

\begin{abstract}
Traditional information security presents a poor ROI: payoffs only manifest when attacks are successfully prevented. In a reality where attacks are inevitable, subpar returns are therefore inevitable. The emerging paradigm of Security Chaos Engineering offers a more remunerative and reliable ROI by minimizing attack impacts and generating valuable evidence to inform continuous improvement of system design and operation.
\end{abstract}

\section{Introduction}
Information security is an economic problem and organizational investment in it therefore bears an expected return. The security program status quo features profligate spending with a slim chance at positive return for the business. Investment is more in feckless ``lemons" than productive ``peaches."

Traditional information security presents a binary success outcome: either the system is ``protected” or else it is ``unprotected.” The return on investment (ROI) hinges on preventing all attacks from occurring. Even if such counterfactuals could somehow be measured in practice, this goal is a fantasy; attacks are inevitable, just like errors and accidents in computer systems and any domain featuring complex systems. Striving for such impossible prevention and the binary of ``secure” or ``insecure” wastes capital that could otherwise be spent on preparing and planning for the inevitable.

The industry’s lack of progress against attackers over decades is widely discussed. The discussion that is overlooked is how security has not made progress in enabling the business to succeed despite the presence of attackers. In fact, tensions between security and other departments, especially software engineering, have deepened over the past decade\cite{b1}.

Because of the binary between ``protected” or ``unprotected,” traditional security is incentivized to implement security measures which obstruct business activity. The implication of this binary goal is that funding a security program at an amount equal or less than the potential cost of compromise – usually framed in terms of a worst-case scenario – is worth the investment. Through this framing, impeding business productivity, growth, and innovation is considered not only reasonable, but a \textit{necessary} negative externality if the organization’s goal is to prevent compromise from happening at all.

There is no current estimate of GDP loss due to \textit{friction} introduced by cybersecurity. It is unclear whether it is more or less than the \textit{potential} GDP loss avoided through successful mitigation, which also lacks direct quantification\cite{b2}. For a true sense of ROI, however, such costs must be taken into consideration. Ironically, this obstruction hinders security efforts as well. Heavy change approval processes slow down an organization’s ability to patch. More generally, inflexibility reduces the ability to adapt to evolving conditions – like attackers shifting towards ransomware to monetize access.

\section{Security Chaos Engineering}
To improve the return on security investment, organizations must embrace a philosophy of resilience: that security controls will fail, mental models of systems will be incomplete, misconfigurations will be overlooked, and attackers will adapt in response to your mitigations — in other words, things will fail and our digital world will constantly evolve.

Security chaos engineering (SCE) represents a new model for security programs grounded in this reality. SCE is the practice of continual experimentation to verify that our systems operate the way we believe and to improve our systems’ resilience to attack\cite{b3}. It extends chaos engineering, which modern enterprises like Netflix already leverage in production to improve system reliability, to the domain of security. SCE, also referred to as ``Continuous Verification," seeks to enable ongoing business success despite the presence of humans who seek to leverage other humans' computers without consent.

\subsection{The ROI of Security Chaos Engineering}
SCE understands that a business which cannot move quickly – whether to innovate, to release changes, to respond to problems – will not be resilient to attacks. Traditional information security calcifies and is insufficiently adaptive to keep up with the evolving nature of the adversarial ``game” being played. Through the framing of SCE, the security program understands and respects that business growth is vital to fund more investment in systems resilience.

In an SCE-driven security program, it is considered a victory to experience the same level of monetary impact from compromises after revenue doubles. If profit margins increase due to productivity boosts, there is also more resource flexibility to adapt to attacks as they evolve over time. SCE's goal is to minimize the suffering of the business in the face of security challenges – whether that is minimizing the impact of attacks or minimizing the friction imposed by security procedures.

Minimizing the impact of attacks, rather than stopping them from happening, is a core driver of SCE's ROI. SCE also helps organizations verify their security investments reflect the potential attack impacts they are prepared to accept in the course of performing digital business activities. That is, SCE exposes any ``lemons" in the organization's security stack. 

\subsection{Aligning Security Mental Models with System Reality}
SCE aligns organizations’ mental models of their system’s security with reality through repeated experimentation. Experiments can detect issues early and proactively by validating hypotheses about system responses to attack scenarios – how the machines and humans constituting and interacting with the system will behave under specific adverse scenarios. An example hypothesis is: ``if a user accidentally or maliciously introduced a misconfigured port, we would immediately detect, block, and alert on the event"\cite{b4}. Without an adequate understanding of how their systems behave under adverse security scenarios, organizations will struggle to know where and how much to invest in security.

The evidence generated by SCE experiments fuels a feedback loop to continuously inform improvements to system design and operation. Observing where and how the system deviated from expectations in an experiment pinpoints how organizations can bolster the system’s resilience to that attack scenario. Experimentation not only allows organizations to catch and fix issues proactively before attackers can take advantage of them, but also verifies whether security controls, procedures, and policies are achieving the security program’s intended outcomes. Detecting faults in security controls proactively can mean the difference between an overlooked misconfiguration and having to announce a data breach to customers. 

Penetration tests take attacker knowledge and apply it one-size-fits all to infrastructure, but few organizations today maintain such cookie-cutter systems. An organization's own engineers are best suited to know what constitutes success and failure states – the system's unique context – and to therefore craft, conduct, and analyze the evidence from SCE experiments to uncover a holistic grasp of system behavior.

\subsection{Incentivizing Resilient System Design}
SCE experiments incentivize teams to consider and plan for failure early, encouraging resilient system design. When combined with experimentation enabled by automation, the result is an evidence-driven feedback loop which informs continuous refinement and cultivates a tangible sense of whether investments are paying off. There is no binary success outcome; a healthy return can come from a variety of wellsprings, whether better containment of attack impact, faster response to and remedy of incidents, or improved engineering productivity and revenue growth without worse incident outcomes.

Making chaos engineering standard practice is correlated with being twice as likely to achieve high resiliency and nearly thrice as likely to ``strongly integrate technology"\cite{b5}. SCE also promotes software velocity, the ability to build and deploy changes on-demand, which is correlated with spending less time remediating security issues\cite{b6}. SCE rejects the notion that business growth is inherently anathema to security; instead, it harnesses and promotes speed. If the organization can deploy end-user feature changes on demand, it can deploy patches on demand – a substantial improvement over the traditional paradigm of unpatched vulnerabilities languishing for months due to concerns about fixes breaking functionality.

\subsection{Building Muscle Memory for Incident Response}
SCE prioritizes the ability to swiftly recover from incidents and harness them as a learning opportunity. By conducting experiments to confirm or deny hypotheses, teams routinely rehearse incident response activities. Each experiment builds ``muscle memory” which transforms stressful firefighting when real attacks occur into confident, practiced problem-solving.

Security chaos experiments also excavate interconnections and interrelations between components – including hardware, software, and humans – as the system operates in production. This richer insight into system dynamics eliminates the instinct to blame ``human error" as the root cause of incidents is shallow and unproductive. When real incidents occur, the muscle memory gained from analyzing experimental evidence discourages knee-jerk explanations in favor of understanding the interactive factors that contributed to the incident.

SCE encourages \textit{learning} from incidents and fosters a more collaborative relationship between security and other teams – an impossibility in the traditional finger-pointing paradigm. The goal outcome is continuous, tangible improvement in the organization's ability to recover gracefully from attacks rather than the traditional security strategy of implementing rigid, punitive policies or controls as ``security theater."

\section{Conclusion}
Information security is currently seen as a cost center, and for good reason. Investments cannot be tied to success outcomes unless all failure is prevented – a folly pursuit. Security chaos engineering, in contrast, offers return on security investment in a variety of dimensions across the software delivery lifecycle. SCE balances reducing incident impact with enabling software velocity and business growth with the understanding that greater organizational flexibility engenders better incident response and resilient design. 

Attackers leverage nimbleness, curiosity, and empiricism to defeat defenders; SCE empowers organizations with the same attributes and ensures resources are spent fruitfully on high-quality, pragmatic peaches rather than futile, profligate lemons.


\begin{thebibliography}{00}

\bibitem{b1} VMWare, Forrester Research, "Bridging the Developer and Security Divide, (2021). [Online]. \url{https://tinyurl.com/2swxp6mj}

\bibitem{b2} Cybersecurity \& Infrastructure Agency, "Cost Of A Cyber Incident," 2020. [Online]. Available: https://tinyurl.com/2p86e56u

\bibitem{b3} Rinehart, A., Shortridge, K. (2020). Security Chaos Engineering. United States: O'Reilly Media, Incorporated.

\bibitem{b4} Rinehart, A., et al. 2017. ChaoSlingr. [Online]. Available:  \url{https://github.com/Optum/ChaoSlingr}

\bibitem{b5} Cisco, "Security Outcomes Study Volume 2," 2021. [Online]. Available: \url{https://tinyurl.com/37aed3hz}

\bibitem{b6} Forsgren, N., Smith, D., Humble, J., \& Frazelle, J. (2019). 2019 Accelerate State of DevOps Report.

\end{thebibliography}
\end{document}